\documentstyle{article}
\newcommand{\be}{\begin{equation}}
\newcommand{\ee}{\end{equation}}
\begin{document}
\begin{titlepage}
\begin{center}
{\Large \bf ORIGIN OF LIGHT SCATTERING \\
FROM DISORDERED SYSTEMS}
\vskip 2 cm
{\it P.Benassi, W.Frizzera, M.Montagna and G.Viliani}\\
Dipartimento di Fisica, Universit\'a di Trento\\
Povo, Trento, I-38050, Italy.
\vskip 0.5 cm
{\it V.Mazzacurati}\\
Dipartimento di Scienze e Tecnologie Biometriche e Biometria\\
Universit\'a dell'Aquila, Collemaggio, L'Aquila, I-67100, Italy.
\vskip 0.5 cm
{\it G.Ruocco and G.Signorelli}\\
Dipartimento di Fisica, Universit\'a dell'Aquila\\
Coppito, L'Aquila, I-67100, Italy.
\end{center}
\end{titlepage}
\newpage

\begin{center}
{\bf ABSTRACT}
\end{center}

Anelastic light scattering is computed numerically for model disordered
systems (linear chains and 2-dimensional site and bond percolators), with
and without electrical disorder. A detailed analysis of the vibrational
modes and of their Raman activity evidences that two extreme mechanisms for
scattering may be singled out. One of these resembles scattering from finite
size systems, while the other mechanisms originates from spatial fluctuations
of the polarizability and is such that modes in even small frequency
intervals may have very different Raman activities. As a consequence,
the average coupling coefficient $C(\omega)$ is the variance of a
zero-average
quantity. Our analysis shows that for both linear chains and 2-dimensional
percolators the second mechanism dominates over the first, and therefore
Raman scattering from disordered systems is essentially due to spatial
fluctuations.

\newpage
\begin{center}
{\bf 1. INTRODUCTION}
\end{center}

Inelastic light scattering is a powerful tool to obtain information on
the vibrational properties of solids. In the case of crystals, the
existence of translational symmetry produces a selection rule which
links the pseudo-momentum of the created or annihilated phonon,
$\overline k$, to the momentum $\overline q$ exchanged between the
photon and the system, i.e.
$\overline k = \overline q$; as
a consequence
acoustic phonons produce Brillouin scattering, while optical phonons at
the centre of the Brillouin zone ($k \ll \pi/a$ where $a$ is
the interatomic distance) produce Raman scattering. The reason why
phonons with $\overline k \neq \overline q$ do not scatter, is that in this
case the electromagnetic waves produced by the polarizability
modulations that are induced at different sites of the crystal by the modes,
interfere in an exactly destructive way.
\par
In disordered solids the situation is quite different.
Brillouin scattering still produces a sharp and intense peak
as in crystals. In fact, the wavelength of the light $\lambda \sim
5000\AA$
is much greater than the correlation length of disorder so that the
vibrational modes with such wavelength are much the same as phonons in
crystals, i.e. plane waves which propagate in the disordered medium
with a
well defined $\overline k$. So the rule $\overline k = \overline q$ is
still
valid. In the case of Brillouin scattering it is customary to speak
about "coherent" scattering
because the polarizability changes induced by the vibrations with $
\overline k = \overline q $ scatter the light inelastically in a
coherent
way, which means that the electric field amplitudes scattered by
atoms placed at different sites sum up in phase. In the following we will
assume $q\simeq 0$ and we will not consider
the Brillouin scattering any more in this paper.
\par
Besides the Brillouin peaks, a continuum spectrum is present; this
is due to the absence of spatial periodicity. In fact, the spatial
fluctuations of the
acousto optical constants prevent the above mentioned completely
destructive interference
among the scattered waves; this produces inelastic scattering which
originates from the spatial variations of the electric polarizability
(disorder-induced light scattering,
DILS) and which has a continuum spectrum ranging from $\omega=0$ up to
some hundred wavenumbers. This is often called "incoherent"
scattering, though in
our opinion this terminology can produce confusion.
\par
In the harmonic approximation the DILS scattered intensity
$I(\omega,T)$
can be put in the form \cite{1}:
$$
I(\omega ) \propto [n(\omega )+1] C(\omega) \rho(\omega) / \omega
$$
where $n(\omega,T)$ is the Bose-Einstein population factor,
$\rho(\omega)$
is the density of vibrational states and $C(\omega)$ is the average
light-vibration coupling coefficient for the modes whose frequency lies
between $\omega$ and $\omega + \delta \omega$:
\be
C_{\alpha \beta}(\omega) =
   {{\sum_p  C_{\alpha \beta}(p) \delta(\omega-\omega_p) } \over
   { \sum_p  \delta(\omega-\omega_p) }}
\label{equa6}
\ee
where
\be
   C_{\alpha \beta}(p) = \bigg{|} \sum_{ij} \sum_{\gamma}
   {{\partial \pi^i_{\alpha \beta}} \over { \partial u^j_\gamma}}
   \bigg{[}
   {{e_\gamma(j|p)}\over{\sqrt{m_j}}} -
   {{e_\gamma(i|p)}\over{\sqrt{m_i}}}
   \bigg{]}
   \bigg{|}^2 = |B_{\alpha\beta}(p)|^2
\label{equa8}
\ee
which defines the quantity $ B_{\alpha\beta}(p) $ whose utility
will become clear later. Note that since $B_{\alpha\beta}(p)$
is a linear function of the atomic displacements
$e(j|p)/\sqrt{m_j}$ produced by mode $p$, and since the sign of the
$e(j|p)$'s
is arbitrary, the only physically relevant quantity is
$|B_{\alpha\beta}(p)|$.
\par
For low frequency acoustic modes (i.e. those whose
wavelength is greater than the correlation length of disorder) what is
expected is $C(\omega) \propto \omega^2$ \cite{2}, which
has been observed experimentally in some systems \cite{3}. At higher
frequency the situation is more complicated.
\par
The DILS Raman scattering is due to a spatially disordered
distribution of the polarizabilities of the scattering units (ions,
atoms or molecules).
In infinite, mechanically and electrically ordered systems the
scattering intensity vanishes. In disordered systems the spatial
polarizability
fluctuations
produce a finite scattering intensity: these fluctuations can be due to
mechanical disorder alone, to electrical disorder alone, or to both.
The situation is somewhat different in finite systems, whose edges can
be
considered as regions where the polarizability varies from the value of
the
scattering units to that of vacuum, so that even fully ordered
{\it finite} systems scatter.
\par
Even though scattering arises in all cases from polarizability
variations,
it may be useful to define two extreme
types of scattering. ($i$) Scattering that is produced by
spatial polarizability
fluctuations, in the sense that different sites have different
polarizabilities. This kind of scattering is present both in
mechanically
ordered \cite{4} and disordered solids. ($ii$) "Finite-size" scattering,
i.e.
the only contribution to scattering present in fully ordered {\it
finite}
systems; this is due to edge effects.
\par
As we will see, even in mechanically disordered systems
some of
the scattering intensity can be ascribed to mechanisms very similar to
edge
effects. We will also see that these two scattering mechanisms have
different characteristics which can be used to define them {\it a
posteriori}.
\par
In the present paper we will study the relative importance of these
two mechanisms; this depends in principle on the scattering system and
on
the frequency range. We will investigate two simple model systems:
linear
chains and two-dimensional site- and bond-percolators: the statistical
method set up for linear chains will be utilized to study the
percolators, which, at threshold, are fractal structures.
\par
Raman scattering from supposedly fractal systems was studied in a
number of recent experimental \cite{bouk} \cite{noiNacl} \cite{tsujimi}
papers. The experimental results seem to imply that the Raman
coupling
coefficient follows a power law $C(\omega) \propto \omega^x$. Much
numerical
work \cite{5}\cite{noiprl}\cite{eric e stoll} was devoted to
determine $x$ in percolators, while theoretical efforts
\cite{alexander} \cite{acw} \cite{duval}
were devoted to derive scaling laws for
$C(\omega)$ in terms of the mechanical parameters (like the fractal
dimension $D$, the spectral dimension $\overline d$, and others)
\cite{alexander} \cite{acw}.
In the next sections, we will show that for the systems under
consideration the DILS mechanism is by far the dominant one.
\par
The paper is organized as follows:
in section 2 we present the results relative to linear chains with
electrical and/or mechanical
disorder. In section 3 we report the results on percolators, while
the final section is devoted to discussion and conclusions.
 \vskip 1 cm
\begin{center}
{\bf 2. LINEAR CHAIN}
\end{center}

In order to better illustrate the scattering mechanisms proposed in the
Introduction, we will apply the previous formalism to various
situations
relative to the linear chain.
\par
Consider an ordered, infinite ($N\rightarrow \infty$) linear chain with
equal masses and equilibrium positions $x^j=aj$, $j=1....N$, connected
by equal springs. In
this case, disregarding Brillouin scattering,
and omitting the polarization subscripts,
equation (\ref{equa8}) becomes:
\be
   C(p) = \bigg{|} \sum_{ij} A_{ij} [ e(j|p) -e(i|p)] \bigg{|}^2
\label{equa9}
\ee
where $A_{ij}=\partial\pi^i/\partial u^j$.
The explicit form of $A_{ij}$
depends on the scattering mechanism used (Dipole-Induced-Dipole (DID),
Bond
Polarizability (BP), etc) but in any case it is an odd function,
$A_{ij} = -A_{ji}$.
\par
It is easy to see that $B(p)=0$ due to cancellation effects
which involve all the scatterers. In
the case of a finite chain, on the contrary,
$B(p) \neq 0$ due to the absence of
complete cancellation for the scatterers at the edges.
  It is clear that ordered systems do
not scatter, and it is only the breakdown of the translational symmetry
of the polarizability which makes finite ordered systems slightly
Raman active.
 \par
 Another interesting case is that
 of electrical disorder in a mechanically ordered system; the coupling
 coefficient was derived in \cite{4}; here we study the distribution
$B(p)$ for one dimensional systems.
Let us consider a linear chain of identical masses
linked by identical next-neighbour springs of rest length $a$, and let
us
assume that the scattering mechanism is next-neighbour DID (ND)
\cite{5}
(extension to three dimensions and full DID (FD) is straightforward).
The electrical disorder is produced by randomly assigning to each mass
a
different isotropic, point-like polarizability $\alpha_i$.
In this case
\[A_{(i-j)}= \left\{ \begin{array}{lll}
			  \alpha^i\alpha^j T^{(3)}(a)  &  \mbox{if i=j+1} \\
			  -\alpha^i\alpha^j T^{(3)}(a) &  \mbox{if i=j-1} \\
			  0                            &  \mbox{otherwise}
			  \end{array}
		\right.  \]
where
\be
     T^{(3)}_{\alpha\beta\gamma}(\bar r) = - \nabla_{\alpha}
\nabla_{\beta}
	\nabla_{\gamma} (1/|\bar r|)
\label{equa16}
\ee
It is easy to obtain, with straightforward but lengthy algebra, the
following expression:
\be
     B(p) = \frac{2}{\sqrt N} T^{(3)}(a) [e^{ik_pa}-1] \bigg{[}
(1+e^{-ik_pa})
     \langle \alpha\rangle  F + G \bigg{]}
\label{equa20}
\ee
where
\be
	F = \sum_j \delta \alpha_j e^{ik_paj}
\label{equa21}
\ee
\be
	G = \sum_j \delta \alpha_j \delta \alpha_{j+1} e^{ik_paj}
\label{equa22}
\ee
For a large number of masses with polarizability  $\alpha_i$ and for many
independent statistical replicas, the statistical distribution of the
quantity $F$
is a gaussian centered at $\langle F\rangle =0$
and with a variance $\sigma^2_F \equiv \langle F^2\rangle  = N \langle
(\delta\alpha)^2\rangle$. The distribution of $G$ is also a
zero-centered gaussian
with $\sigma^2_G = (\sigma^2_F)^2/N$. Therefore the distribution of
$B(p)$,
which is the convolution of two gaussians, is again a zero centered
gaussian with a variance $\sigma^2_B$ given by:
\begin{eqnarray}
\nonumber
    \sigma^2_B = \bigg{|} {{2T^{(3)}(a)} \over {N}} (e^{ik_pa}-1)
		 \bigg{|}^2 [(1+e^{-ik_pa})\langle \alpha\rangle \sigma^2_F
+\sigma^2_G] \\
		 = {{9}\over{4N}}  \bigg{(} {{2} \over {a}} \bigg{)}^8
		[  \langle \alpha\rangle ^2 \langle (\delta \alpha)^2\rangle
sin^2(k_pa) + \langle (\delta
	 \alpha)^2\rangle ^2 sin^2(k_pa/2)]
\label{equa23}
\end{eqnarray}
in agreement with eqs. (22), (31), (35) and (36) of reference \cite{4}.
\par
{}From this analysis it is clear that the origin of DILS is in fluctuations:
in fact, $C(p)$ is just the variance of the zero-average quantity
$B(p)$.
 \par
We consider next a linear chain with $m_1=1$ and $m_2=2$
randomly distributed with equal probability and linked by identical
springs
of elastic constant $K=1$, with periodic boundary conditions.
The maximum frequency is $\omega_{max}=2\sqrt{K/\mu}$ where $\mu$ is
the lighter mass, and in the following $\omega$ will be given in units
of $\omega_{max}$. In the
infinite chain all modes would be localized \cite{7}, but in a finite
chain
the localization length of the lowest energy modes may be larger than
the
chain length, so that these modes appear not to be localized. Roughly
speaking, there are two kinds of modes:
\begin{itemize}
\item acoustic-like  modes;
\item very localized modes,
\end{itemize}
We assign the polarizabilities $\alpha_1, \alpha_2$ to the masses $m_1$
and
$m_2$ respectively.
Vibrational
eigenvalues and eigenvectors were obtained by diagonalizing the
dynamical
matrix of many replicas of the system. In Fig. 1(a) we report $D(|B|)$,
i.e.
the distribution
of the $|B(p)|$ relative to the 2 modes with lowest frequency of 2000
chains containing 150 masses each, with $\alpha_1=0  $
$\alpha_2=2$. $C(\omega)$
may
be then obtained from equation (\ref{equa6}) by averaging on the modes
with energy $\omega \pm \delta
\omega$ which are produced by the different realizations.
\par
In order to test if $D(|B|)$ is a gaussian, we computed its first 10
moments:
$$
    M_n = \int dB D(|B|) |B|^n
$$
and compared them with the corresponding ones, $M^G_n$, of the gaussian
having
the same zero and second moment. The ratios $M_n/M^G_n$ are reported in
Table I together with the $\chi ^2$ test values.
We find that the gaussian distribution is
a very good approximation for all acoustic-like modes (i.e. for $\omega
< 0.5 $), so that the introduction of mechanical disorder does not change
the results we obtained for electrical disorder alone.
For some frequency intervals for
$\omega > 0.5$ important departures from gaussian shape are observed which
show up in the
appearance of peaks at well defined $|B|$ values (see Fig. 1(b)).
\par
By choosing narrower frequency intervals on which the averages
are taken it is possible to enhance the
peaks (see Fig. 1(c)).
The frequencies around which Fig. 1(a) and Fig. 1(b-c) were
computed are indicated by the arrows in Fig. 2, where $C(\omega)$ is
shown.
The modes which produce the peaks at $|B| \neq 0$ are those where two
neighboring light masses are surrounded by two long sequences of heavy
masses, and oscillate in counter phase with $\omega \approx 0.9$, while
the heavy masses are practically stationary.
These modes scatter by a mechanism analogous to that of finite size
systems: they cannot produce exact cancellation of the scattering
amplitude.
 \par
The present analysis evidences therefore that there are two extreme
mechanisms
for Raman activity in disordered linear chains: DILS and  finite-size
scattering. The former produces a distribution $D(|B|)$ which is a
gaussian
centered at zero, while the latter produces peaks.
In any case we have seen that the DILS mechanism is the largely dominant
one when many masses are involved in the mode.
Only in very narrow and high frequency ranges one can find modes
localized on few atoms which give finite-size scattering.

\begin{center}
{\bf 3. SCATTERING FROM 2-DIMENSIONAL PERCOLATING STRUCTURES}
\end{center}
The same analysis as in the previous section was performed on
2-dimensional
site and bond percolators at percolation threshold, containing
identical
masses with identical bare polarizability. In the systems, each mass is
bound to its nearest  neighbours, the connecting springs are identical
and
scalar elasticity is assumed. Periodic boundary conditions are imposed
and
the eigenvalues are normalized to the maximum frequency of a full
square lattice $\omega_{max}=2\sqrt{2K/ \mu}$.
The Raman coupling coefficient was computed assuming that the
scattering mechanism is full DID, nearest neighbour DID, and Bond
Polarizability (BP) \cite{9}.
\par
In two dimensions equation (\ref{equa8}) for FD becomes
\cite{5}:
\be
C_{\alpha\beta}(p)=
\bigg{|} 2\alpha^2\sum_{ij}\sum_{\gamma}
T^{(3)}_{\alpha\beta\gamma}(\overline{r}^{ij})
[e_{\gamma}(j|p) - e_{\gamma}(i|p)]\bigg{|}^2=
|B_{\alpha \beta}(p)|^2
\label{equa24}
\ee
To calculate the depolarized Raman scattering, $B_{xy}(p)$ (or
$B_{yx}(p)$)
are used, while for the polarized scattering $B_{xx}(p)$, $B_{xy}(p)$ and
$B_{yy}(p)$ are used.
We have computed
separately the statistical distributions for the two groups of $B$'s
($B_{\alpha\alpha}$ or $B_{\alpha\beta}$ with $\alpha\neq\beta$).
In Fig. 3 (a-c) we report the $B_{\alpha\alpha}$ distributions
relative to the modes
in three frequency intervals (0.08-0.12, 0.37-0.43, 0.57-0.63, respectively)
for 450 replicas of a 20$\times$20
site percolator. For the two lowest-frequency intervals the
distributions
are very well fitted by  gaussians, as shown in Table I. For the
frequency range  $0.57 < \omega < 0.63$, important
deviations from the gaussian shape are observed near the origin, and
are quantified in Table I. In particular, there is an excess in the
first channel, showing that many modes are little Raman active, which
means that cancellation effects are very relevant.
By expanding the abscissa (Fig. 4) it comes out that most of the excess
modes are
confined to a very small range of $|B|$ values, of the order of
$\approx 10^{-5}$ times the value of $ \sigma$ of
the distribution of Fig. 3 (c).
\par
In order to gain more insight on the nature of these low activity
modes, we tried to characterize their wavefunction by the localization
length evaluated according to the Thouless definition:
$$
	    l_p = R_p^{-1/D}
$$
where $D$ is the fractal dimension and $R$ is the participation ratio:
\be
		R_p= \frac {\sum_i [e(i|p)]^4}{(\sum_i [e(i|p)]^2)^2}
\label{equa25}
\ee
In Fig. 5(a) we report the distribution of $l$ for the $\approx$ 10$^4$
fractons with
frequency $0.57 <\omega < 0.63$, whose $D(|B|)$ is shown in Fig. 3(c).
As can be seen, in this frequency range fractons with very different
localization lengths are present, in qualitative agreement with
previous
calculations \cite{10} \cite{11}. If, from the modes which produce Fig.
3(c), we select those with $|B|<10^{-3}$
and consider the distribution of their localization lengths, we obtain
Fig. 5(b). Most of these low-activity modes are localized on a few
masses, and
all have practically the same frequency ($\omega \approx 0.61$).
Examples of this localized, low Raman active modes are in Fig. 6 (a-b)
which shows that only few masses are involved in the mode (4 and 8
in Fig. 6(a) and 6(b) respectively), while all the other masses are
practically stationary.
In the same frequency range ($\omega \approx 0.61$) however, there are
many more modes (see Fig. 5(a))
involving many masses,  as exemplified in Fig. 6(c).
\par
In order to check if a correlation between Raman activity and
localization length exists, we also studied the $D(|B|)$ relative to groups
of modes having about the same $l$. Exception made for modes
with $l<3$ (which are those that produce the peak in $B=0$ of Fig. 3(c))
all the others groups produce gaussian distributions for $D(|B|)$
with exactly the same variance and, therefore, the same $C(\omega)$.
This result was to be expected as a consequence of DILS: in fact,
within the bond polarizability model, $B(p)$ is the sum
of $n$ fluctuating terms ($n\approx l^D$ is the number of
participating bonds),
each having a mean squared amplitude $\approx 1/n$. This is true if, for a
mass involved  in mode $p$, $\langle(u_i-u_j)^2\rangle \propto \langle u^2
\rangle$, because  $\langle u^2 \rangle \propto 1/n$ for normalized modes.
Thus in this case the quantity $\langle B^2 \rangle$
does not depend on $n$ or $l$,  giving rise to the same $C(\omega)$ for
modes with very different localization length.
\par
Full DID in site percolators was studied as well. What we
observe in this case is that the longer range of the FD interaction
tends to smooth out the extremes of almost zero Raman activity
observed above, so
that it is less simple to single out the very localized modes, and the
distributions look more like gaussians.
\par
Bond percolators were also studied. $50\times 50$ samples have been
used to study the low frequency range (Fig. 7(a)).
In this case and for FD, as Figs. 7 (b-c) show, there are some
statistically significant departures (see Table I) from
the gaussian shape at high frequencies, though the
distributions still appear to be peaked
at zero. This result was expected because in this case  there may be
disconnected pairs of nearest neighbours masses and since connected
and disconnected nearest neighbours pairs give different DID
contributions to $|B|$, a single gaussian distribution is
unlikely to reproduce $D(|B|)$.
On the contrary, BP produces gaussian-like distributions in
this case also.
\begin{center}
{\bf 4. CONCLUSIONS}
\end{center}
In the present work we studied electrically disordered linear chains
with
and without mechanical disorder, and two dimensional site and bond
percolators without electrical disorder.
\par
For the mechanically ordered linear chains, it was analytically shown
that
the distribution $D(|B|)$ obtained for each mode by randomly assigning
the
bare polarizability to the masses, is a gaussian centered at zero.
Therefore, the Raman coupling coefficient of the modes is the variance
of
the distribution. By introducing mechanical disorder, no relevant
difference is observed at low frequencies, while at higher frequencies
$D(|B|)$ shows peaks centered either at zero or at finite values of
$|B|$. The latter peaks are due to
 very localized, Raman active modes. Thus, this study singled out two
scattering mechanisms: DILS (which is intrinsically due to
spatial fluctuations)
and  finite-size scattering.
\par
The same statistical analysis was performed on site and bond
percolators,
and what emerges is that the DILS mechanism dominates in all
frequency ranges and for both scattering mechanisms, FD and BP.
\par
The fluctuation origin of Raman scattering in these systems is
shown by the Gaussian shape of the distribution $D(|B|)$.

\newpage

\begin{center}
FIGURE CAPTIONS
\end{center}
Fig. 1. Distribution of the $|B|$ values relative to 2000
replicas of mechanically and electrically disordered chains containing 150
masses each.
(a) $m_1=1$, $m_2=2$, $\alpha_1=0$ and $\alpha_2=2$ , the first
two modes $0.005 <\omega <0.018$.
(b) $m_1=1$, $m_2=2$, $\alpha_1=0.9$, $\alpha_2=1.1$, for all
the eigenvalues falling in the frequency range $0.8994 <\omega <0.9050$.
(c) Same as (b), but for $0.899450 <\omega <0.899475$.
\par
Fig. 2. $C(\omega)$ for electrically and mechanically
disordered linear chains. The arrows indicate the frequencies for which
the statistical analysis relative to figures 1(a-b) was performed.
\par
Fig. 3. Distribution of the $|B_{\alpha\alpha}|$ values computed in the
bond polarizability model for all modes of 450 replicas
of a
20$\times$20 site percolator. (a): $0.08 <\omega <0.12$; (b): $0.37
<\omega <0.43$; (c): $0.57 <\omega <0.63  $. The arrow in (c) shows the
number of modes which are left in the first channel  neglecting those with
$|B_{xx}|$ and $|B_{yy}|$ $ <10^{-3}$.
\par
Fig. 4. Same as Fig. 3(c), but plotted for $|B_{\alpha\alpha}|<10^{-3}$.
\par
Fig. 5. (a) Distribution of the localization lengths of $\approx$ 10$^4$
fractons
with frequency  $0.57 <\omega <0.63  $, i.e. the same which produce
Fig. 3(c).
(b) Same as (a), but considering only the fractons with
$|B_{xx}|$ and $|B_{yy}|$ $ <10^{-3}$.
\par
Fig. 6. Modes of a $20\times 20$ site percolator with $\omega \approx 0.61$.
The localization length are: (a) $l \approx 2.1$,
(b) $l \approx 3.0$, (c) $l \approx 7.5$.
\par
Fig. 7. Distribution of $|B_{\alpha\alpha}|$ values for modes of a bond
percolator  (FD).
(a) $50 \times 50$ percolator in the range $0.014 < \omega < 0.018$.
(b) $20 \times 20$ percolator in the range $0.37 < \omega < 0.43$
(c) $20 \times 20$ percolator in the range $0.57 < \omega < 0.63$
\vskip 1 cm
\begin{center}
TABLE CAPTION
\end{center}
Table I. Ratios between the first 10 moments, $M_n$, of the distribution
$D(|B|)$ and the moments of a gaussian having the same zero
and second moment of $D(|B|)$. The values of the reduced $\chi ^2$
are also reported.

\newpage

\begin{table}
\caption [TABLE I]
\bigskip
\begin{center}
\begin{tabular}{|c|c|c|c|c|c|c|c|} \hline
Figures & 1(a) & 3(a) & 3(b) & 3(c) & 7(a) & 7(b) & 7(c)    \\
\hline
$\chi ^2$     & 1.08 & 1.32 & 1.04 & 3.87 & 1.18 & 7.84 & 11.8\\
M$_1$/M$^G_1$ & 1.00 & 1.01 & 1.00 & 1.00 & 0.99 & 0.97 & 0.97\\
M$_2$/M$^G_2$ & 1.00 & 1.00 & 1.00 & 1.00 & 1.00 & 1.00 & 1.00\\
M$_3$/M$^G_3$ & 1.00 & 0.97 & 1.00 & 0.99 & 1.03 & 1.07 & 1.07\\
M$_4$/M$^G_4$ & 1.01 & 0.94 & 1.00 & 0.97 & 1.06 & 1.19 & 1.18\\
M$_5$/M$^G_5$ & 1.02 & 0.89 & 0.99 & 0.93 & 1.11 & 1.36 & 1.33\\
M$_6$/M$^G_6$ & 1.04 & 0.85 & 0.99 & 0.89 & 1.16 & 1.58 & 1.53\\
M$_7$/M$^G_7$ & 1.06 & 0.80 & 0.99 & 0.84 & 1.20 & 1.86 & 1.79\\
M$_8$/M$^G_8$ & 1.09 & 0.75 & 0.99 & 0.78 & 1.24 & 2.21 & 2.11\\
M$_9$/M$^G_9$ & 1.13 & 0.70 & 1.00 & 0.72 & 1.25 & 2.62 & 2.50\\
M$_{10}$/M$^G_{10}$ & 1.16 & 0.65 & 1.00 & 0.65 & 1.25 & 3.08 & 2.00\\
\hline
\end{tabular}
\end{center}
\end{table}

\begin{thebibliography}{99}
\bibitem{9} In the case of site percolators, nearest neighbour DID and
Bond
Polarizability are the same mechanism, see Ref. \cite{5}.
\end{thebibliography}
\end{document}